\documentclass[conference]{IEEEtran}
\IEEEoverridecommandlockouts
\usepackage{cite}
\usepackage{amsmath,amssymb,amsfonts}
\usepackage{bbm}
\usepackage{algorithm,algorithmic}
\usepackage{graphicx}
\usepackage{textcomp}
\usepackage{xcolor}
\usepackage{float}
\usepackage{amsmath}
\usepackage{caption}
 \usepackage{geometry}
\newtheorem{theorem}{Theorem}
\newtheorem{lemma}{Lemma}

\graphicspath{{figures/}}
\allowdisplaybreaks
\begin{document}

\title{Joint Link Rate Selection and Channel State Change Detection in Block-Fading Channels}
\author{
	\IEEEauthorblockN{Haoyue~Tang\textsuperscript{1,*},~Xinyu~Hou\textsuperscript{1,*},~Jintao~Wang\textsuperscript{1,2},~Jian~Song\textsuperscript{1,2}}
	
	\IEEEauthorblockA{
		\textsuperscript{1}Beijing National Research Center for Information Science and Technology (BNRist),\\
		Dept. of Electronic Engineering, Tsinghua University, Beijing 100084, China\\
		\textsuperscript{2}Research Institute of Tsinghua University in Shenzhen, Shenzhen, 518057\\
		\{thy17@mails,houxy19@mails,  wangjintao@, jsong@\}tsinghua.edu.cn
}}

\maketitle

\begin{abstract}
 In this work, we consider the problem of transmission rate selection for a discrete time point-to-point block fading wireless communication link. The wireless channel remains constant within the channel coherence time but can change rapidly across blocks. The goal is to design a link rate selection strategy that can identify the best transmission rate quickly and adaptively in quasi-static channels. This problem can be cast into the stochastic bandit framework, and the unawareness of time-stamps where channel changes necessitates running change-point detection simultaneously with stochastic bandit algorithms to improve adaptivity. We present a joint channel change-point detection and link rate selection algorithm based on Thompson Sampling (CD-TS) and show it can achieve a sublinear regret with respect to the number of time steps $T$ when the channel coherence time is larger than a threshold. We then improve the CD-TS algorithm by considering the fact that higher transmission rate has higher packet-loss probability. Finally, we validate the performance of the proposed algorithms through numerical simulations. 
\end{abstract}


\section{Introduction}
\let\thefootnote\relax\footnotetext{\noindent -----------------\\ * The two authors contributes equally. 
\\This work was supported by the National Key R\&D Program of China (Grant No.2017YFE0112300). Corresponding author: Jintao Wang}

Selecting a proper transmission rate that adapts channel conditions is important efficient communications in the current wireless systems, i.e., devices using IEEE802.11 standards \cite{80211_tmc,combs_jsac}. Considering the fact that precise channel state information (CSI) feedback is supported by few 802.11 devices \cite{feedback} but ACK/NACKs of the transmission outcomes are always available to the transmitter, rate sampling approaches \cite{ra_1,ra_2} that select transmission rate based on historical acknowledgments have been widely used. 
Online sequential decision making frameworks such as stochastic bandit provide efficient solutions to optimize rate sampling methods \cite{gupta_ts_18infocom} when the channel statistics remains constant. However, the use of high frequency spectrum (e.g., mmWave) in the 5G wireless communication systems \cite{fcc} causes CSI to change more frequently than in conventional spectrums. 
This motivates us to incorporate change-point detection algorithms into rate sampling link rate selection strategy in order to improve the cumulative link throughput.


Employing stochastic bandit algorithms for link rate selection in slow and fast fading channels have been widely studied previously. Gupta \emph{et al.} designed a link rate selection algorithm based on Thompson Sampling (TS) and showed the proposed algorithm achieves an average expected regret of $\mathcal{O}(\log T)$ \cite{gupta_ts_18infocom}. The algorithm is then applied to the joint selection of modulation and transmission rate \cite{qi_80211_cl}. 
By utilizing the correlations between transmission success probabilities and transmission rates, \cite{gupta_19_infocom,muhammed_infocom20, tong_2020_icc} designed modified TS algorithms that can further decrease the expected cumulative throughput regret in higher confidence intervals. The aforementioned techniques are useful under the assumption that channel statistics remain constant, which is true for both slow are fast fading scenarios. However, when channels experience abrupt changes in block fading scenarios and the channel statistics become non-stationary, those algorithms may suffer from slow adaptation. Although TS with a sliding window is proposed in \cite{qi_80211_cl} to overcome this issue and is shown to achieve a high cumulative throughput empirically, theoretic evaluations on regret performance are not well addressed. 

Designing fast adaptive stochastic bandit algorithms in quasi-static environments are studied in \cite{cao_19_cdbandit,liu_bandit_aaai}. It is shown that by running a simple CD algorithm to control the clearance of historical data, the expected regret of Upper Confidence Bound (UCB) algorithm in the quasi-static environment can be improved. However, these algorithms are designed for general bandit problems and can be improved when applied to link rate selection scenarios. 

Employing change-point detection algorithms to design efficient transmission schemes for block-fading and quasi-static channels has recently been considered in \cite{wu_19_cdglobecom}. The key is to estimate channel change points accurately from the high dimensional received signals, and then re-estimate the new CSI and re-design beamforming strategies for channel throughput maximization. 
Notice that their algorithm relies on the sparse structure of the received signal in angular domain subspace, which requires the receiver to possess large-scale uniform antenna arrays. Therefore, the algorithm cannot be applied directly to mobile devices with a small number of antennas.


To overcome the aforementioned challenges, in this work, we investigate the problem of link rate selection in block fading quasi-static channels. The distribution of channel change points and the probabilities of successful transmission are unknown. The goal is to design a transmission strategy that attempts to maximize the cumulative channel throughput by using historical transmission outcomes. We present an efficient change-point and online decision making algorithm, and then derive the corresponding expected regret upper bound. We then improve the empirical performance of the algorithm by constraining the TS posteriors to lay in a set that can characterize the relationship between transmission success probabilities and transmission rates.

The rest of the paper is organized as follows: in Section II we introduce the system model and formulate the overall optimization problem. Section III proposes a joint channel change-point detection and Thompson Sampling based link rate selection algorithm (CD-TS). We then provide theoretic analysis of the expected regret upper bound. Section IV improves the proposed CD-TS by constraining the sampling posteriors. Section V validates the proposed algorithms via numerical simulations and Section VI draws the conclusion. 

\section{Problem Formulation}
We consider a communication rate selection problem for a discrete-time point-to-point wireless channel and let $t\in\{1, \cdots, T\}$ denote the index of the current slot. The communication channel between the transmitter and the receiver is quantized into $Q<\infty$ states, and let $h(t)\in\mathcal{Q}=\{1, \cdots, Q\}$ denote the current quantized channel state. 
The channel $h(t)$ experiences block fading and during slot $[0, T]$, there exists $M(M\geq 1)$ channel state change points, whose timestamps are denoted by $\{\nu_1, \nu_2, \cdots, \nu_M\}$. The channel state $h(t)$ remains the same during the interval between two change points, i.e., $h(t_1)=h(t_2), \forall t_1, t_2\in[\nu_m, \nu_{m+1}-1]$ but changes rapidly before and after the change point, i.e., $h(\nu_m-1)\neq h(\nu_m), \forall m$. The difference between two nearby change points is lower bounded by the channel coherence time $L$, i.e.,  $\nu_{i+1}-\nu_i\geq L$, where $L\gg 1$ is predetermined by the physical channel and mobility characteristics. 

In each slot $t$, the transmitter chooses a transmission rate $r_{i(t)}\in \mathcal{R}=\{r_1, \cdots, r_R\}$ indexed by $i(t)$. 
The wireless channel is erroneous and let $X(t)\in\{0, 1\}$ be the indicator function of whether the transmission in slot $t$ succeeds: if $X(t)=1$, all of the $r_{i(t)}$ transmitted packets will be successfully received by the end of slot $t$; otherwise, all  $r_{i(t)}$ packets will be lost due to decoding error \cite{gupta_19_infocom,gupta_ts_18infocom}. The transmission success probabilities depend on the current transmission and channel state, since using higher transmission rate will lead to higher packet-loss probabilities. Therefore, we assume that the transmission success probability follows a Bernoulli distribution with parameter $\mathbb{E}[X(t)]=\theta_{h(t), i(t)}$. With no loss of generality,  assume that transmission rates are arranged in an increasing order, i.e., $r_1<\cdots< r_R$, and therefore $1>\theta_{q, 1}>\cdots>\theta_{q, R}>0$. The distribution of $X(t)$ is independent of transmission outcome of other slots $X(t'), \forall t'\neq t$. At the end of slot $t$, the receiver sends $X(t)$ to the transmitter without error. 

The goal of this research is to maximize the cumulative throughput by designing a link rate selection strategy $\pi$ that chooses $i(t)$ based on historical transmissions. The problem is organized as follows:

\emph{Problem:}
\begin{align}
	\max_{\pi}&\sum_{t=1}^Tr_{i(t)}\theta_{h(t), i(t)}, \nonumber\\
	\text{~where~}&\pi:\{((i(1), X(1)), \cdots, (i(t-1), X(t-1))\}\rightarrow i(t).\label{eq:obj}
\end{align}


\section{Change-Point Detection aided Thompson Sampling Algorithm (CD-TS)}

Maximizing \eqref{eq:obj} can be cast in the stochastic bandit framework. However, traditional bandit algorithms assume the distribution of reward remains constant across all the slots. Therefore, adopting these algorithms directly to the quasi-static channel link rate selection may cause large regret, because historical data cannot reflect the current best rate after the channel changes. To overcome this issue, 
we propose to run a change-point detection (CD) algorithm simultaneously with stochastic bandit to control the clearance of historic data and therefore, improve adaptivity in block-fading channels. 

Assume $\hat{\mathcal{T}}=\{\hat{\tau}_1, \hat{\tau}_2, \cdots\}$ is the detected channel change points sequence and let $c(t)=\max_{m}\{\hat{\tau}_m|\hat{\tau}_m\leq t\}$ denote the time-stamp of the most recently detected change point before slot $t$. We use $N_i(t), s_i(t)$ and $f_i(t)$ to denote the recorded number of total transmissions, success and failure times using transmission rate $i$ from $c(t)+1$ to $t-1$. Link rate selection algorithm for slot $t$ based on $N_i(t), s_i(t), f_i(t)$ is introduced in Section III-A. Channel change points detection algorithms to calculate $\{\hat{\tau}_m\}$ are proposed in Section III-B. The two parts of the algorithm are synthesized in Section III-C. 

\subsection{Thompson Sampling for link rate selection}

Consider that in slot $t$, selecting rate $i$ has succeeded for $s_i(t)$ and failed for $f_i(t)$ times from slot $c(t)+1$ to $t-1$. If the channel has not changed after $c(t)$, the posterior distribution of transmission success probability $\lambda_i=\theta_{h(t), i}$ follows a Beta distribution parameterized by $s_i(t)$ and $f_i(t)$. The probability density function of such Beta distribution is:
\begin{align}
    p_i(\lambda)=&\text{Beta}(s_i(t)+1, f_i(t)+1)\nonumber\\
    =&\frac{1}{Z_i(s_i(t), f_i(t))}\lambda^{s_i(t)}(1-\lambda)^{f_i(t)},
\end{align}
where $Z_i(x, y)=\int_{0}^1\lambda^x(1-\lambda)^y\mathsf{d}\lambda$ is a normalizing constant. 



The Thompson Sampling algorithm for link rate selection is as follows: first sample $\hat{\lambda}_i\sim\text{Beta}(s_i(t)+1, f_i(t)+1)$ and then choose rate $i(t)=\arg\max_{i}\hat{\lambda}_ir_i$ for transmission. 




\subsection{Change-Point Detection}

Recall that CSI changes lead to variations in transmission success probabilities of every transmission rate. Therefore, we attempt to detect past channel changes by comparing the empirical transmission success probability of the latest $w$ transmission outcomes and the second latest $w$ transmission outcomes of a transmission rate. We require there is at least a specific transmission rate indexed by $i_{\text{cd}}$ that is selected every $F$ slots, so that the change-point detection algorithm can always be run when needed. In this work, we select $i_{\text{cd}}$ to be the rate with the highest historical empirical throughput during slot $c(t)+1\sim c(t)+F$. 

To efficiently implement the change-point detection, we use $W_i(k)$ to record the $k$-th transmission outcome of using rate $i$. If rate $i(t)$ has been selected more than $2w$ times, i.e., $N_i(t)>2w$, we run a CD algorithm by comparing the mean of the latest $w$ outcomes of using rate $i(t)$ (denoted by $\mathbb{M}_1$) and the second latest $w$ transmission outcomes (denoted by $\mathbb{M}_2$), where $\mathbb{M}_1$ and $\mathbb{M}_2$ can be computed by:
\begin{subequations}
\begin{align}
    &\mathbb{M}_1:=\frac{1}{w}\sum_{k=1}^{w}W_{i(t)}(N_i(t)-w+k), \\
    &\mathbb{M}_2:=\frac{1}{w}\sum_{k=1}^{w}W_{i(t)}(N_i(t)-2w+k).
\end{align}\label{eq:mean}
\end{subequations}

If the absolute difference between the two mean values is larger than a predetermined threshold $b$, i.e., $|\mathbb{M}_2-\mathbb{M}_1|>b$, we conclude that a channel change point occurs recently  and append $t$ to the detected change points set $\hat{\mathcal{T}}$.

        
\subsection{Algorithm Design}
We then synthesize our proposed algorithm CD-TS by combining the aforementioned CD and TS parts together.

First we initialize the historical transmission records $N_i(0)=0, s_i(0)=0, f_i(t)=0, \forall i\in\mathcal{R}$ and set $c(1)=0$. Then in each slot $t$, we perform a joint link rate selection and channel change-point detection as follows:

\begin{itemize}

	\item \textbf{Transmission Rate Selection:}
	
	If $t=c(t)+kF, \forall k\in\mathbb{N}^+$, then we should transmit with rate indexed by $i(t)=i_{\text{cd}}=\arg\max_i \frac{s_i(c(t)+F)}{N_i(c(t)+F)}r_i$ for efficient channel change-point detection;
	
	Otherwise, we select transmission rate based on Thompson Sampling. We sample $\hat{\lambda}_i\sim\text{Beta}(s_i(t)+1, f_i(t)+1)$ and treat it as an estimation of transmission success probability using rate $i$. Then we transmit with $i(t)=\arg\max_i\hat{\lambda}_ir_i$, observe the transmission outcome $X(t)$ and update $N_i(t+1)=N_i(t)+1, s_i(t+1)=s_i(t)+X(t),  f_i(t+1)=f_i(t)+(1-X(t))$. 
	
	\item \textbf{Channel Change-Point Detection:} 
	If $N_i(t)>2w$, we run a change-point detection by computing the mean transmission outcomes $\mathbb{M}_1$ and $\mathbb{M}_2$ using \eqref{eq:mean}. If $|\mathbb{M}_1-\mathbb{M}_2|>b$, we conclude there is a change point before $t$ and record by $c(t+1)=t$. Then we clear historical transmission records by setting $N_i(t+1), s_i(t+1), f_i(t+1)$ to $0$. 
\end{itemize}

The algorithm flow chart is provided in the flow chart \ref{alg}. 
\begin{algorithm}
	\caption{Joint channel change detection and Thompson Sampling}\label{alg:CD-CoTS}
	\textbf{Initialization: }For each rate index $i$, set $N_i(t), s_i(t), f_i(t)\leftarrow 0$. Let the most recently detected channel change be $c(1)=0$. 
	
	\begin{algorithmic}
		\FOR{$t=1, 2, \cdots T$}
		\IF{$t=c(t)+kF, k\in\mathbb{N}^+$} 
		
		\STATE Set $i(t)\leftarrow \max_i \frac{s_i(c(t)+F)}{N_i(c(t)+F)}r_i$.  \algorithmiccomment{Select $i_{\text{cd}}$ for CD}
		
		\ELSE 
		
		\STATE {Sample rate $\hat{\lambda}_i\sim\text{Beta}(s_i(t)+1, f_i(t)+1)$. } 
		\STATE Select rate $i(t)\leftarrow\arg\max_{i}\hat{\lambda}_ir_i$. \algorithmiccomment{TS}
		
		\ENDIF
		
		\STATE Transmit at $i(t)$, observe ACK/NACK $X(t)$ and record $W_{i(t)}(N_i(t)+1)=X(t)$.
		
		\IF{$N_i(t)+1>2w$}
		\STATE Compute mean success probability $\mathbb{M}_1, \mathbb{M}_2$ from $\{W_i(k)\}$ using \eqref{eq:mean}.
		\IF{$\left|\mathbb{M}_1-\mathbb{M}_2\right|>b$}
		\STATE \algorithmiccomment{Change detected, clear historical data}
		
		\STATE $c(t+1)\leftarrow t, N_i(t+1), s_i(t+1), f_i(t+1)\leftarrow 0, \forall i$.
		
		\ELSE 
		\STATE \algorithmiccomment{Record historical data}
		
		\STATE $c(t+1)\leftarrow c(t), N_{i(t)}(t+1)\leftarrow N_{i(t)}(t)+1, s_{i(t)}(t+1)\leftarrow s_{i(t)}(t)+X(t), f_{i(t)}(t+1)\leftarrow f_{i(t)}(t)+(1-X(t))$. 
		
		\ENDIF
		\ENDIF
		
		\ENDFOR
	\end{algorithmic}
\label{alg}
\end{algorithm}

\subsection{Regret Analysis}
We evaluate the performance of the proposed CD-TS algorithm via the cumulative expected throughput regret against the optimum link rate selection algorithm when the precise CSI is known. 
Denote $i^*(h)\triangleq\arg\max_i r_i\theta_{h, i}$ as the index of optimal rate when the channel state is $h$. 
By definition, the cumulative regret in non-stationary environment denoted by $\mathcal{R}_T$, is the difference between the expected cumulative throughput of using $\{i(t)\}_{t=1}^T$ chosen by our algorithm and the optimum rate using $\{i^*(h(t))\}_{t=1}^T$, i.e.,
\begin{align}
	\mathcal{R}_T&\triangleq\mathbb{E}\left[\sum_{t=1}^T\left(r_{i^*(h(t))}\theta_{h(t), i^*(h(t))}-r_{i(t)}\theta_{h(t), i(t)}\right)\right]. 
	\label{eq:regretdef}
\end{align}

A lower regret \eqref{eq:regretdef} implies a higher expected cumulative throughput and indicates the algorithm adapts faster to the block fading channel. To compute the expected regret of the proposed algorithm, we first introduce the following lemma, the proof is provided in Appendix~\ref{proof-lemm1}:

\begin{lemma}\label{lemma1}
Consider the channel remains stable during slot $1\sim T$, i.e., $h(t)=h$. Let $\Delta_i=r_{i^*(h)}\theta_{h, i^*(h)}-r_i\theta_{h, i}$ be the expected throughput difference between choosing rate $i$ and the optimum rate $i^*(h)$, the expected regret of the proposed algorithm for the this time invariant channel, denoted by $\tilde{\mathcal{R}}_T$ can be upper bounded by the following inequality for any $\epsilon\in(0, 1]$: 
\begin{align}
    \tilde{\mathcal{R}}_T\leq&
    \max_i\Delta_i\left(2T^2\exp\left(-\frac{wb^2}{2}\right)+\frac{T}{F}\right)\nonumber\\&+B\log \left(1-\frac{1}{F}\right)T+\mathcal{O}\left(\frac{R}{\epsilon^2}\right),
\end{align}
where 
coefficient $B$ is a constant unrelated to $T$,
\[B=(1+\epsilon)\sum_{i\neq i^*(h)}\frac{\mathbbm{1}\left(\frac{r_{i^*(h)}\theta_{h, i^*(h)}}{r_i}<1\right)}{D_{\text{KL}}\left(\theta_i, \frac{r_{i^*(h)}\theta_{h, i^*(h)}}{r_i}\right)}\Delta_i,\]
and $D_{\text{KL}}(p, q)=p\log\frac{p}{q}+(1-p)\log\frac{1-p}{1-q}$ is the KL divergence. 
\end{lemma}

With Lemma 1, we proceed to our main theorem, whose proof is provided in \ref{pf:thm1}:
\begin{theorem}
Suppose $\delta_{\text{min}}=\min_{h, h', i}|\theta_{h, i}-\theta_{h', i}|$, if there exists $\alpha>0$ so that $L\geq\frac{48}{\delta_{\text{min}}^2}(MT)^{\alpha}\log MT$, then the regret of the proposed CD-TS scales with $\mathcal{O}((MT)^{\max(\alpha, 1-\alpha)}\log MT)$ if $w$,  $b$ and $F$ are set to $b=\delta_{\text{min}}/2$, $w=\frac{12}{\delta_{\text{min}}^2}\log MT$ and $F=(MT)^\alpha$. 
\end{theorem}

\section{Improving Cumulative Throughput with Constrained Thompson Sampling}
In this section, we improve the performance of CD-TS by replacing TS with Constrained Thompson Sampling (CoTS) algorithm proposed in \cite{gupta_19_infocom}. CoTS can achieve a lower regret performance by utilizing the fact that choosing higher transmission rate will have lower success probability. 
Recall that $\theta_{h,i}$ is the transmission success probability if the channel state $h(t)=h$ and $r(t)=r_i$. To exploit the relationship between $\{\theta_{h, i}\}_{i=1}^N$ for better exploration, we first define set $\Theta$, where the element of each vector $\boldsymbol{\lambda}\in\Theta$ is monotonically decreasing:
\begin{equation}
\Theta:=\{\boldsymbol{\lambda}\in\Theta|\theta_{i_1}>\theta_{i_2}, \forall i_1<i_2\}.     
\end{equation}
Thus, for any fixed channel state $h(t)=h$, the transmission success probability $\boldsymbol{\theta}_{h, :}$ should belong to set $\Theta$ due to the fact that transmitting at higher rate is more likely to get packet-losses. 
Then we sample $\hat{\boldsymbol{\lambda}}(t)=[\hat{\lambda}_1, \cdots, \hat{\lambda}_N]$ jointly from:
\begin{equation}\hat{\boldsymbol{\lambda}}\sim\frac{1}{Z}\mathbbm{1}(\boldsymbol{\lambda}\in\Theta)\prod\text{Beta}(s_i(t)+1, f_i(t)+1),
\end{equation}
where $Z$ is a normalizing constant.

To implement CoTS during link rate selection, we first sample $\hat{\lambda}_i\sim\text{Beta}(s_i(t)+1, f_i(t)+1)$ independently for each rate. If $\hat{\boldsymbol{\lambda}}\in\Theta$ we stop and proceed to select the rate $i(t)=\arg\max_{i}\hat{\lambda}_ir_i$ for transmission. Otherwise we keep sampling  $\hat{\lambda}_i\sim\text{Beta}(s_i(t)+1, f_i(t)+1)$ until $\hat{\boldsymbol{\lambda}}\in\Theta$.

\section{Simulation Results}
In this section, we provide numerical simulations to validate the performance gain of our proposed algorithms. We consider link rate selection settings in IEEE802.11a/g systems \cite{6242359}, where we can choose from 8 possible transmission rates from set $\mathcal{R}=[6, 9, 12, 18, 24, 36, 48, 54]$ (in Mbps) \cite{gupta_ts_18infocom}. Suppose the channel is quantized into $3$ states and the transmission success probabilities are shown in TABLE \ref{states}. The cumulative regret is computed for $T=$3000 slots with coherence time $L\geq 750$ and $M=3$ change points, and the time-stamps of the channel change points are marked in the figure. The expected regret is computed by taking the average of 100 runs. 

\begin{table}[h]
\caption{Channel state settings}
\centering
\begin{tabular}[h]{c|c}
\hline
Channel States & Transmit Success Possibility                  \\ \hline
state1         & [0.59,0.45,0.34,0.22,0.15,0.10,0.03,0.01] \\
state2         & [0.79,0.74,0.65,0.63,0.52,0.35,0.26,0.22] \\
state3         & [0.99,0.95,0.90,0.85,0.80,0.76,0.60,0.52] \\
\hline
\end{tabular}
\label{states}
\end{table}

\begin{figure}[h]
\centering
\includegraphics[width=.4\textwidth]{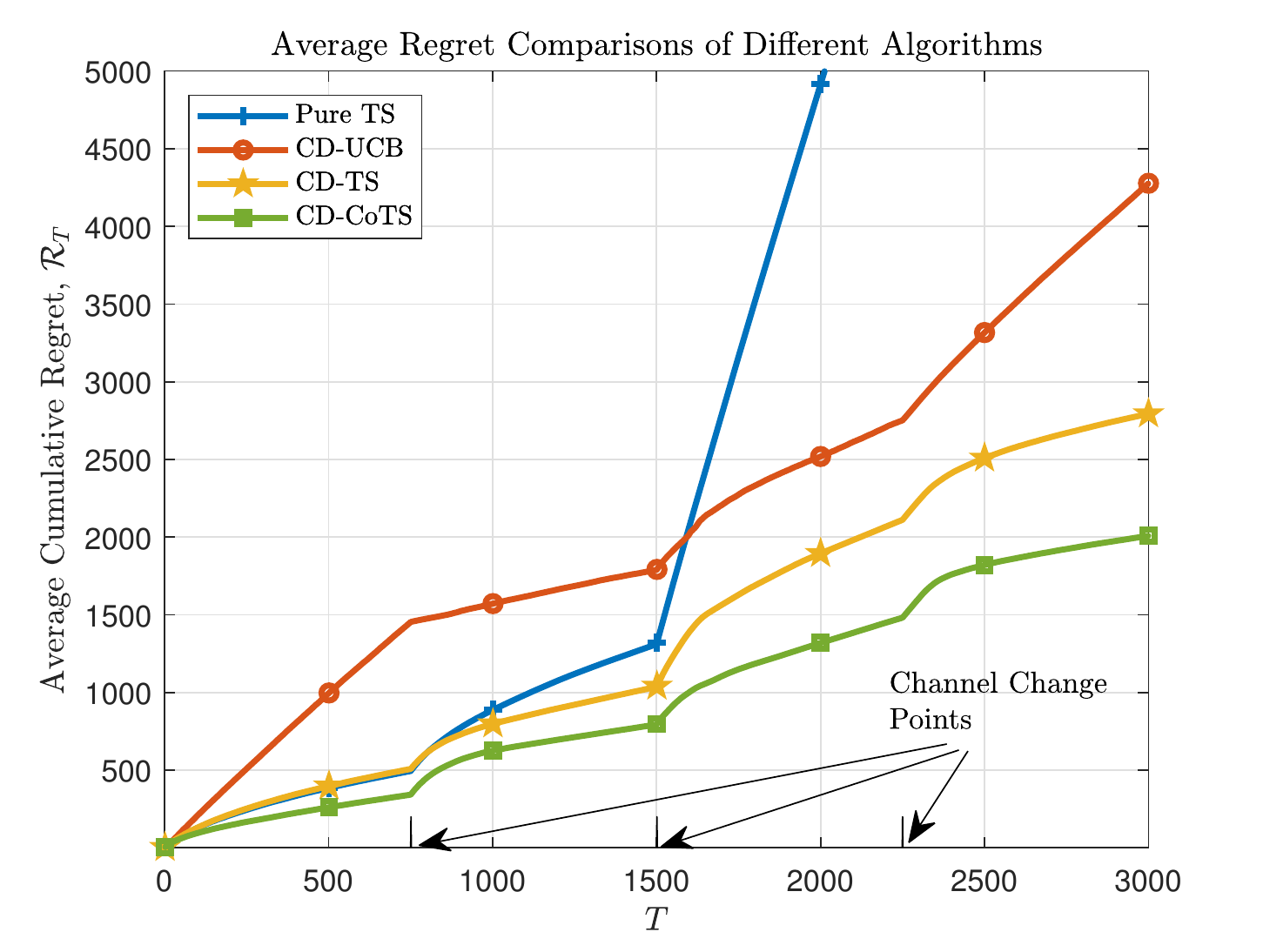}
\caption{Cumulative Regret Comparisons of Various Algorithms}
\label{fig:regret}
\end{figure}

\begin{figure}[h]
\centering
\includegraphics[width=.4\textwidth]{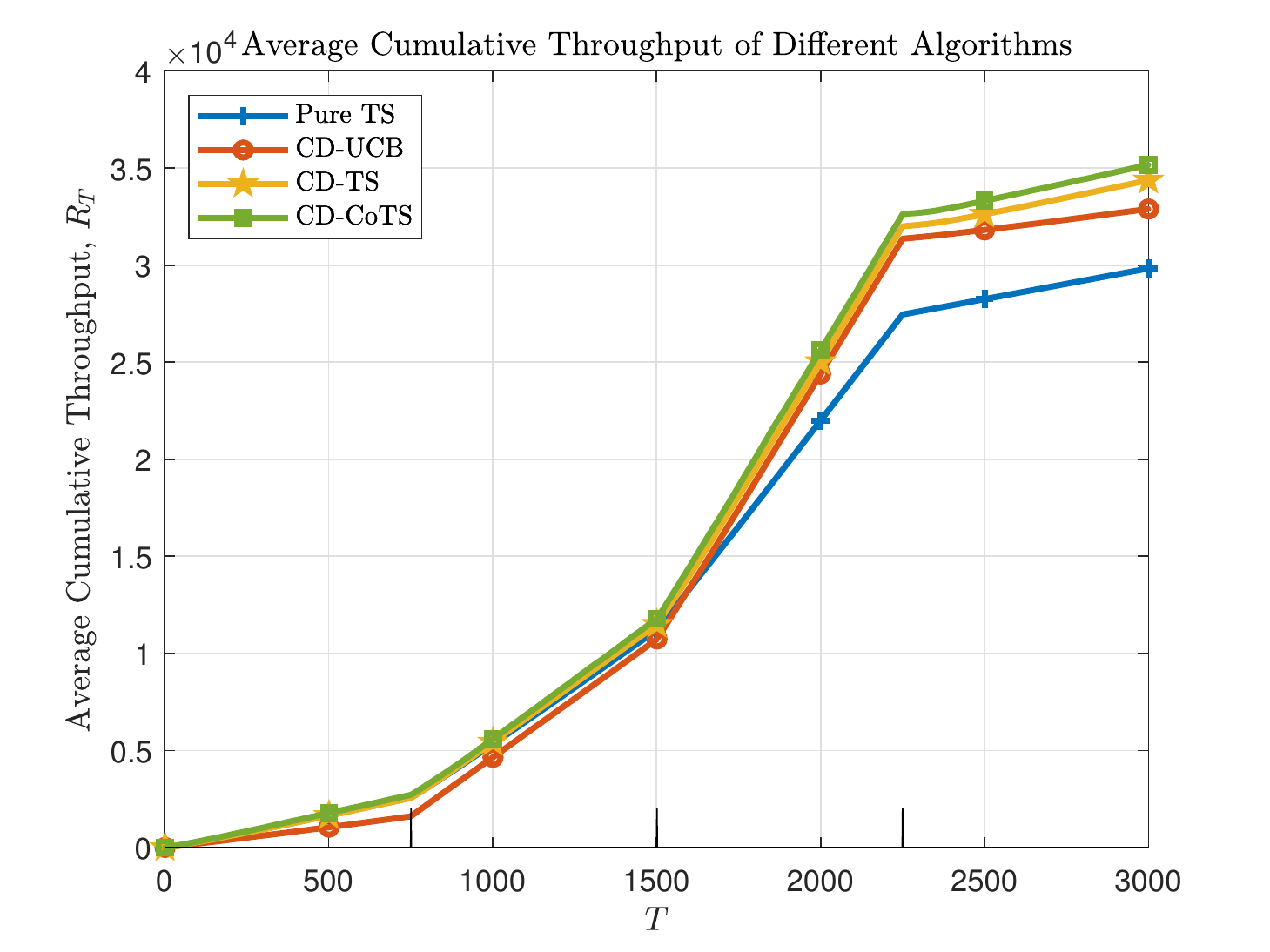}
\caption{Average Cumulative Throughput Comparisons of Different Algorithms}
\label{fig:rate}
\end{figure}

Fig.~\ref{fig:regret} plots the expected cumulative regret and Fig.~\ref{fig:rate} plots the cumulative throughput of the proposed CD-TS, CD-CoTS,  CD-UCB \cite{cao_19_cdbandit} and pure Thompson Sampling \cite{gupta_19_infocom} algorithm with no change-point detection. The expected throughput/regret are computed by taking the average over 100 runs. From the Fig.~\ref{fig:regret}, the proposed CD-TS and CD-CoTS achieve small regret by incorporating change-point detection into stochastic bandit framework. This is because when the channel changes, the proposed CD-TS and CD-CoTS algorithms can efficiently detect the changes, and they can fit quickly to newly changed channel state by throwing away historical data, while the pure TS algorithm without change-point detection algorithm does not clear up historical data and therefore experiences linear regret once the channel changes. 
Compared with CD-UCB algorithm, the proposed CD-TS and CD-CoTS algorithms can further reduce the cumulative throughput regret. 


The decrease in cumulative regret implies that the proposed CD-TS and CD-CoTS algorithms can achieve higher cumulative throughput as depicted in Fig.~\ref{fig:rate}. Compared with the CD-TS algorithm, CD-CoTS algorithm achieves a smaller cumulative regret and higher cumulative throughput empirically by taking the non-increasing characteristics of transmission success probabilities into account. 
\section{Conclusions}
In this paper, we study rate sampling based link rate selection problem in block fading channel. We propose a CD-TS algorithm that detects abrupt channel change points by using a sliding window and then choose the best transmission rate based on Thompson Sampling (CD-TS). We show that, theoretically, the proposed CD-TS can achieve a sub-linear regret with respect to the total number of time steps when the channel coherence time is above a certain threshold. We then proceed to improve CD-TS by sampling packet-loss probabilities of different rates from a constrained posterior set, in which higher transmission rates have higher packet-loss probabilities. Simulation results show that, by detecting channel change points and performing adaptive link rate selection jointly,  the proposed CD-TS achieves a smaller cumulative regret and higher cumulative throughput. The proposed CD-CoTS algorithm further improves CD-TS empirically by taking the distribution of transmission success probabilities into account. 

Future work includes theoretic analysis of the CD-CoTS algorithm and extending both CD-TS/CD-CoTS to multi-users scenario with block fading channels. Moreover, currently the selection of threshold $b$ and window size $w$ depend on time length $T$ and the number of change points $M$. It will be of interest to design algorithms that can achieve sub-linear regret without knowledge of $T$ and $M$. 
\appendices
\section{Proof of Lemma 1}\label{proof-lemm1}
For time invariant channel with fixed channel state $h$, let $\omega$ be a sample path and $\hat{\mathcal{T}}=\{\hat{\tau}_1(\omega), \cdots\}$ are the time slots of detected channel change points on the sample path. For simplicity, denote $i^*=i^*(h)=\arg\max_i r_i\theta_{h, i}$ to be the optimum transmission rate. Let $\tilde{R}_{T}(\omega)$ be the cumulative expected throughput regret of sample path $\omega$ for the time-invariant channel, i.e., 
\[\tilde{R}_{T}(\omega)\triangleq\sum_{t=1}^{T}(r_{i^*}\theta_{h, i^*}-r_{i(t;\omega)}\theta_{h, i(t;\omega)}).\]

Since there is no channel change points before slot $T$, event $\mathcal{F}=\{\omega:\hat{\tau}_1(\omega)<T\}$ denotes the CD algorithm raises at least one false alarm before $T$. By conditional expectation, we can decompose $\mathcal{R}_T$ into:
\begin{align}
    \tilde{\mathcal{R}}_T=\text{Pr}(\mathcal{F})\mathbb{E}[\tilde{R}_T(\omega)|\mathcal{F}]+\text{Pr}(\mathcal{F}^c)\mathbb{E}[\tilde{R}_T(\omega)|\mathcal{F}^c].\label{eq:append1regret}
\end{align}


We proceed to bound each item on the RHS respectively. To upper bound the first item $\text{Pr}(\mathcal{F})\mathbb{E}[\tilde{R}_T(\omega)|\mathcal{F}]$, we introduce Lemma~\ref{lemma2} to upper bound the probability of false alarm $p_{\text{FA}}$ in each slot. The proof of which is a direct application of Hoeffding bound and is omitted due to space limitations. 
\begin{lemma}\label{lemma2}
In each slot $t$, the probability that a false alarm raises can be upper bounded by:
\begin{equation}
    p_{\text{FA}}\leq 2\exp\left(-\frac{b^2w}{2}\right).
\end{equation}
\end{lemma}


Since we can run the change-point detection algorithm for at most $T$ times from slot $1$ to $T$, by union bound we can upper bound $\text{Pr}(\mathcal{F})$ by:
\begin{equation}
    \text{Pr}(\mathcal{F})\leq T p_\text{FA}.
\end{equation}



Plugging the above equations into the first term on the RHS of \eqref{eq:append1regret}, we have:
\begin{equation}
    \text{RHS}_1\leq T p_{\text{FA}}\mathbb{E}[\max_i\Delta_i\hat{\nu}_1|\mathcal{F}]\leq \max_i\Delta_iT^2p_{\text{FA}}
\end{equation}

Next we proceed to upper bound the second item $\text{Pr}(\mathcal{F}^c)\mathbb{E}[\tilde{R}_T(\omega)|\mathcal{F}^c]$. To do this, let policy $\hat{\pi}$ be algorithm without a change-point detection, i.e., in each slot select $i(t)$ based on Thompson Sampling and transmit with rate $i_{\text{cd}}$ every $F$ slots. Notice that for $\omega\in\mathcal{F}^c$, transmission strategy obtained by the proposed CD-TS algorithm and $\hat{\pi}$ is exactly the same. Therefore:
\begin{align}
    &\text{Pr}(\mathcal{F}^c)\mathbb{E}_{\text{CD-TS}}\left[\tilde{R}_T(\omega)|\mathcal{F}^c\right]\nonumber\\
    =&\text{Pr}(\mathcal{F}^c)\mathbb{E}_{\hat{\pi}}\left[\tilde{R}_T(\omega)|\mathcal{F}^c\right]\nonumber\\
    \overset{(a)}{\leq} &\text{Pr}(\mathcal{F}^c)\mathbb{E}_{\hat{\pi}}\left[\tilde{R}_T(\omega)|\mathcal{F}^c\right]+\text{Pr}(\mathcal{F})\mathbb{E}_{\hat{\pi}}\left[\tilde{R}_T(\omega)|\mathcal{F}\right]\nonumber\\
    =&\mathbb{E}_{\hat{\pi}}\left[\tilde{R}_T(\omega)\right]\nonumber\\
    =&\mathbb{E}_{\hat{\pi}}\left[\sum_{i=1, i\neq kF}^T(r_{i^*}\theta_{h, i^*}-r_{i(t, \omega)}\theta_{h, i(t;\omega)})\right]\nonumber\\
    &+\mathbb{E}_{\hat{\pi}}\left[\sum_{i=1, i= kF}^T(r_{i^*}\theta_{h, i^*}-r_{i(t, \omega)}\theta_{h, i(t;\omega)})\right],\label{append1:RHS2}
\end{align}
where (a) is obtained because regret is non-negative. Then according to \cite[Theorem 1]{gupta_ts_18infocom}, the regret caused by TS $\tilde{\mathcal{R}}_T^{\text{TS}}:=\mathbb{E}_{\hat{\pi}}\left[\sum_{i=1, i\neq kF}^T(r_{i^*}\theta_{h, i^*}-r_{i(t, \omega)}\theta_{h, i(t;\omega)})\right]$ can be upper bounded by:
\begin{equation}
    \tilde{\mathcal{R}}_T^{\text{TS}}\leq B\log \left(1-\frac{1}{F}\right)T+\mathcal{O}\left(\frac{R}{\epsilon^2}\right),
\end{equation}
where \[B=(1+\epsilon)\sum_{i\neq i^*(h)}\frac{\mathbbm{1}\left(\frac{r_{i^*(h)}\theta_{h, i^*(h)}}{r_i}<1\right)}{D_{\text{KL}}\left(\theta_i, \frac{r_{i^*(h)}\theta_{h, i^*(h)}}{r_i}\right)}\Delta_i,\]
and $D_{\text{KL}}(p, q)=p\log\frac{p}{q}+(1-p)\log\frac{1-p}{1-q}$, $\Delta_i=r_{i^*(h)}\theta_{h, i^*(h)}-r_i\theta_{h, i}$ are the KL divergence and difference in expected throughput, respectively. The regret caused by frequent sample $i_{cd}$ can be simply upper bounded by: \begin{align}\tilde{\mathcal{R}}_T^{\text{FS}}:=\mathbb{E}_{\hat{\pi}}\left[\sum_{i=kF}(r_{i^*}\theta_{h, i^*}-r_{i(t, \omega)}\theta_{h, i(t;\omega)})\right]\!\leq\! \max_i\Delta_i\lfloor\frac{T}{F}\rfloor.\nonumber
\end{align}
Plugging the above items into \eqref{append1:RHS2} and then into \eqref{eq:append1regret} yields Lemma 1. 

\section{Proof of Theorem 1}\label{pf:thm1}
The change-point detection algorithm guarantees after the latest change point has been detected for $2wF$ slots, rate $i_{\text{cd}}$ can be selected at least $2w$ times for runnning change point detection algorithm. Denote $\mathcal{M}_{m}:=\{\omega:\hat{\tau}_m>v_m+2wF\}$ to be the event that the proposed algorithm does not detect change point $m$ after $2wF$ slots. 
Then by properties of conditional expectation, the regret can be upper bounded by:
\begin{align}
    \mathcal{R}_{T}=&\text{Pr}(\mathcal{M}_1^c)\mathbb{E}[R_T(\omega)|\mathcal{M}_1^c]+\text{Pr}(\mathcal{M}_1)\mathbb{E}[R_T(\omega)|\mathcal{M}_1]\nonumber\\
    \overset{}{=}&\text{Pr}(\mathcal{M}_1^c)\mathbb{E}[R_{1:\nu_1}(\omega)|\mathcal{M}_1^c]+\text{Pr}(\mathcal{M}_1^c)\mathbb{E}[R_{\nu_1+1:\hat{\tau}_1}(\omega)|\mathcal{M}_1^c]\nonumber\\
    &+\text{Pr}(\mathcal{M}_1^c)\mathbb{E}[R_{\hat{\tau}_1+1:T}(\omega)|\mathcal{M}_1^c]\nonumber\\
    &+\text{Pr}(\mathcal{M}_1)\mathbb{E}[R_{1:\nu_1}(\omega)|\mathcal{M}_1]+\text{Pr}(\mathcal{M}_1)\mathbb{E}[R_{\nu_1+1:T}(\omega)|\mathcal{M}_1]\nonumber\\
    \overset{(a)}{\leq} &\tilde{\mathcal{R}}_{\nu_1}+\max_i\Delta_i\times 2wF+\mathbb{E}[R_{\hat{\tau}_1+1:T}(\omega)|\mathcal{M}_1^c]\nonumber\\
    &+\max_{i}\Delta_i\text{Pr}(\mathcal{M}_1)T,
    \label{thm1:eq1}
\end{align}
where equality (a) is because $\tilde{\mathcal{R}}_{\nu_1}=\text{Pr}(\mathcal{M}_1^c)\mathbb{E}[R_{1:\nu_1}(\omega)|\mathcal{M}_1^c]+\text{Pr}(\mathcal{M}_1)\mathbb{E}[R_{1:\nu_1}(\omega)|\mathcal{M}_1]$, and if $\mathcal{M}_1^c$ happens, the detection delay $\hat{\tau}_1-\nu_1\leq 2wF$. 

To upper bound $\text{Pr}(\mathcal{M}_1)$, we then introduce Lemma~\ref{lemma3}, which upper bounds the probability of miss detection for any channel state $h(t)=h$:
\begin{lemma}\label{lemma3}
Let $\delta_{\text{min}}=\min_{h, h', i}|\theta_{h, i}-\theta_{h', i}|$, then probability $\text{Pr}(\mathcal{M}_m)$ can be upper bounded by:
\begin{equation}
    \text{Pr}(\mathcal{M}_m)\leq p_{\textsuperscript{MD}}\leq 2\exp\left(-\frac{w(b-\delta_{\text{min}})^2}{2}\right),
\end{equation}
where $\delta_{\text{min}}\triangleq\min_{h, h', i}|\theta_{h, i}-\theta_{h', i}|$ is the minimum absolute difference of transmission success probability.
\end{lemma}
The proof is provided in Appendix~\ref{pf:lemma3}

Let $T_m=\nu_{m}-\nu_{m-1}$ be the duration of the $m$-th stationary period. Notice that for each $\hat{\tau}_1$, the expected regret $\mathbb{E}[R_{\hat{\tau}_1+1:T}(\omega)|\mathcal{M}_1^c]=\mathcal{R}_{T-\hat{\tau_1}}$. Repeating the manipulation in \eqref{thm1:eq1} for $\mathcal{R}_{T-\hat{\tau}_1}$ and finally we have:
\begin{align}
    \mathcal{R}_T\leq&2MT\max_i\Delta_i\exp\left(-\frac{w(\delta_{\text{min}}-b)^2}{2}\right)+\sum_{m=1}^M\tilde{\mathcal{R}}_{T_m}\nonumber\\
    &+\max_i\Delta_i\times 2wFM.
\end{align}

Plugging $\tilde{\mathcal{R}}_{T_m}$ from Lemma~\ref{lemma1} into the equation, we have:
\begin{align}
    \mathcal{R}_T\leq&\max_i\Delta_i\times M\left(2T\exp\left(-\frac{w(\delta_{\text{min}}-b)^2}{2}\right)\right.\nonumber\\
    &\left.+2T^2\exp\left(-\frac{wb^2}{2}\right)+\frac{T}{F}+2wF\right)\nonumber\\
    &+B\left(\sum_{m=1}^M\log T_m\right).
    \label{eq:append2regret}
\end{align}

Next, we discuss parameter tuning problems to achieve a sub-linear expected regret. We simply set $b=\frac{1}{2}\delta_{\text{min}}$ for easy implementations. To guarantee $2MT\exp\left(-\frac{w(\delta_{\text{min}}-b)^2}{2}\right)\leq C_1$ and $2MT^2\exp\left(-\frac{wb^2}{2}\right)\leq C_2$, we require $w$ to satisfy:
\begin{align}
w\geq&\frac{1}{\delta_{\text{min}}^2/8}\max\{\log\frac{2MT}{C_1}, \log\frac{2MT^2}{C_2}\}. 
\end{align} 

For simplicity, we choose $C_2=2\sqrt{MT}$ and let $w=\frac{12}{\delta_{\text{min}}^2}\log MT$ so that the first and second item on the RHS of \eqref{eq:append2regret} are sub-linear with respect to $T$. Achieving a total sub-linear regret $\tilde{R}_T$ then requires $MT/F$ and $MwF$ to be sub-linear. To achieve this, we set $F=(MT)^{\alpha}(\alpha>0)$ so that $MT/F$ is sublinear to $T$. If the channel coherence time $L$ satisfies:
$
	L/2\geq 2wF=\frac{24}{\delta_{\text{min}}^2}(MT)^{\alpha}\log MT$,
then the proposed algorithm achieves a sub-linear expected regret of $\mathcal{O}((MT)^{\max(\alpha, 1-\alpha)}\log T)$.

 \section{Proof of Lemma~\ref{lemma2}}\label{pf:lemma2}
 Our change point detection compares the mean $\mathbb{M}_1$ and $\mathbb{M}_2$ computed from \eqref{eq:mean}. If event $|\mathbb{M}_1-\mathbb{M}_2|>b$ happens, then at least one of the events $\left|\mathbb{M}_1-\theta_{i_{cd}, h}\right|>\frac{b}{2}$, $\left|\mathbb{M}_2-\theta_{h, i_{cd}}\right|>\frac{b}{2}$ may happen. By using the union bound we have:

\begin{align}
     P_{\text{FA}}&=\text{Pr}(|\mathbb{M}_1-\mathbb{M}_2|>b)\nonumber\\
     &\leq \text{Pr}(|\mathbb{M}_1-\theta_{h, i_{cd}}|>b/2)+\text{Pr}(|\mathbb{M}_2-\theta_{h, i_{cd}}|>b/2)\nonumber\\
     &\leq 2\exp\left(-\frac{wb^2}{2}\right), 
 \end{align}
 where the last inequality is obtained by Hoeffding inequality. 

 \section{Proof of Lemma~\ref{lemma3}}\label{pf:lemma3}
 Suppose the channel in the $m$-th stationary period is $h$ and denote $\delta=|\theta_{h, i_{cd}}-\theta_{h', i_{cd}}|$. If event $\mathcal{M}_m$ happens, i.e., $|\mathbb{M}_1-\mathbb{M}_2|>b$. Therefore, since change point $m$ is not detected, it can be conducted that either event $\left|\mathbb{M}_1-\theta_{i_{cd}, h}\right|>\frac{\delta-b}{2}$ or event $\left|\mathbb{M}_2-\theta_{h, i_{cd}}\right|>\frac{\delta-b}{2}$ must happen. By using the union bound and Hoeffding inequality we have:
 \begin{align}
     &\text{Pr}(\mathcal{M}_m)\nonumber\\
     \leq& \text{Pr}(|\mathbb{M}_1-\theta_{h, i_{cd}}|>\frac{\delta-b}{2})+\text{Pr}(|\mathbb{M}_2-\theta_{h', i_{cd}}|>\frac{\delta-b}{2})\nonumber\\
     \leq & 2\exp\left(-\frac{w(b-\delta)^2}{2}\right).\label{eq:lemm3-last}
 \end{align}
 Recall that $\delta_{\text{min}}=\min_{h, h', i}|\theta_{h, i}-\theta_{h', i}|$ be the minimum absolute difference of transmission success probability, by taking the minimum on both sides of \eqref{eq:lemm3-last}, we have:
 \begin{equation}
     \text{Pr}(\mathcal{M}_m)\leq p_{\text{MD}}\leq 2\exp\left(-\frac{w(b-\delta_{\text{min}})^2}{2}\right). 
 \end{equation}

\bibliography{bibfile}

\begin{thebibliography}{10}
\providecommand{\url}[1]{#1}
\csname url@samestyle\endcsname
\providecommand{\newblock}{\relax}
\providecommand{\bibinfo}[2]{#2}
\providecommand{\BIBentrySTDinterwordspacing}{\spaceskip=0pt\relax}
\providecommand{\BIBentryALTinterwordstretchfactor}{4}
\providecommand{\BIBentryALTinterwordspacing}{\spaceskip=\fontdimen2\font plus
\BIBentryALTinterwordstretchfactor\fontdimen3\font minus
  \fontdimen4\font\relax}
\providecommand{\BIBforeignlanguage}[2]{{%
\expandafter\ifx\csname l@#1\endcsname\relax
\typeout{** WARNING: IEEEtran.bst: No hyphenation pattern has been}%
\typeout{** loaded for the language `#1'. Using the pattern for}%
\typeout{** the default language instead.}%
\else
\language=\csname l@#1\endcsname
\fi
#2}}
\providecommand{\BIBdecl}{\relax}
\BIBdecl

\bibitem{80211_tmc}
R.~Combes, J.~Ok, A.~Proutiere, D.~Yun, and Y.~Yi, ``Optimal rate sampling in
  802.11 systems: Theory, design, and implementation,'' \emph{IEEE Transactions
  on Mobile Computing}, vol.~18, no.~5, pp. 1145--1158, 2019.

\bibitem{combs_jsac}
R.~Combes and A.~Proutiere, ``Dynamic rate and channel selection in cognitive
  radio systems,'' \emph{IEEE Journal on Selected Areas in Communications},
  vol.~33, no.~5, pp. 910--921, 2015.

\bibitem{feedback}
D.~J. Love, R.~W. Heath, V.~K. N.~Lau, D.~Gesbert, B.~D. Rao, and M.~Andrews,
  ``An overview of limited feedback in wireless communication systems,''
  \emph{IEEE Journal on Selected Areas in Communications}, vol.~26, no.~8, pp.
  1341--1365, 2008.

\bibitem{ra_1}
A.~Kamerman and L.~Monteban, ``Wavelan®-ii: A high-performance wireless lan
  for the unlicensed band,'' \emph{Bell Labs Technical Journal}, vol.~2, no.~3,
  pp. 118--133, 1997.

\bibitem{ra_2}
J.~Bicket, ``Bit-rate selection in wireless networks,'' 09 2006.

\bibitem{gupta_ts_18infocom}
H.~Gupta, A.~Eryilmaz, and R.~Srikant, ``Low-complexity, low-regret link rate
  selection in rapidly-varying wireless channels,'' in \emph{IEEE INFOCOM 2018
  - IEEE Conference on Computer Communications}, 2018, pp. 540--548.

\bibitem{fcc}
``Fcc adopts rules to facilitate next generation wireless technologies,''
  \emph{Federal Communications Commission, Tech Report}, July 2016.

\bibitem{qi_80211_cl}
H.~Qi, Z.~Hu, X.~Wen, and Z.~Lu, ``Rate adaptation with thompson sampling in
  802.11ac wlan,'' \emph{IEEE Communications Letters}, vol.~23, no.~10, pp.
  1888--1892, 2019.

\bibitem{gupta_19_infocom}
H.~{Gupta}, A.~{Eryilmaz}, and R.~{Srikant}, ``Link rate selection using
  constrained thompson sampling,'' in \emph{IEEE INFOCOM 2019 - IEEE Conference
  on Computer Communications}, 2019, pp. 739--747.

\bibitem{muhammed_infocom20}
M.~A. Qureshi and C.~Tekin, ``Online bayesian learning for rate selection in
  millimeter wave cognitive radio networks,'' in \emph{IEEE INFOCOM 2020 - IEEE
  Conference on Computer Communications}, 2020, pp. 1449--1458.

\bibitem{tong_2020_icc}
J.~Tong, S.~Lai, L.~Fu, and Z.~Han, ``Optimal frequency and rate selection
  using unimodal objective based thompson sampling algorithm,'' in \emph{ICC
  2020 - 2020 IEEE International Conference on Communications (ICC)}, 2020, pp.
  1--7.

\bibitem{cao_19_cdbandit}
Y.~Cao, Z.~Wen, B.~Kveton, and Y.~Xie, ``Nearly optimal adaptive procedure with
  change detection for piecewise-stationary bandit,'' in \emph{Proceedings of
  the Twenty-Second International Conference on Artificial Intelligence and
  Statistics (AISTATS2019)}, vol.~89, 16--18 Apr 2019, pp. 418--427.

\bibitem{liu_bandit_aaai}
F.~Liu, J.~Lee, and N.~B. Shroff, ``A change-detection based framework for
  piecewise-stationary multi-armed bandit problem,'' in \emph{Proceedings of
  the Thirty-Second {AAAI} Conference on Artificial Intelligence, (AAAI-18),
  New Orleans, Louisiana, USA, February 2-7, 2018}.\hskip 1em plus 0.5em minus
  0.4em\relax {AAAI} Press, 2018, pp. 3651--3658.

\bibitem{wu_19_cdglobecom}
Y.~Wu, Y.~Jiao, F.~Gao, and Y.~Gu, ``Pilot-free channel change detection for
  mmwave massive mimo system,'' in \emph{2019 IEEE Global Communications
  Conference (GLOBECOM)}, 2019, pp. 1--6.

\bibitem{6242359}
T.~{Lin}, C.~{Tsai}, and K.~{Wu}, ``Earc: Enhanced adaptation of link rate and
  contention window for ieee 802.11 multi-rate wireless networks,'' \emph{IEEE
  Transactions on Communications}, vol.~60, no.~9, pp. 2623--2634, 2012.

\end{thebibliography}

\end{document}